# Hydrogen jet diffusion modeling by using physics-informed graph neural network and sparsely-distributed sensor data

Xinqi Zhang[1], Jihao Shi[1,2*], Junjie Li[1], Xinyan Huang[2], Fu Xiao[2], Qiliang Wang[2], Asif Sohail Usmani[2], Guoming Chen[1]

[1] Centre for Offshore Engineering and Safety Technology, China University of Petroleum, Qingdao 266580, China

[2] Department of Building Environment and Energy Engineering, The Hong Kong Polytechnic University, Kowloon, Hong Kong, China

*Corresponding author

Email address: shi_jihao@163.com, jihao.shi@polyu.edu.hk

**Abstract:**

Transitioning to hydrogen as a clean energy carrier is a promising climate mitigation strategy. However, hydrogen's propensity to leak poses explosion risks if ignited. Efficient modeling of jet diffusion during accidental release is critical for operation and maintenance management of hydrogen facilities. Deep learning has proven effective for concentration prediction in gas jet diffusion scenarios. Nonetheless, its reliance on extensive simulations as training data and its potential disregard for physical laws limit its applicability to unseen accidental scenarios. Recently, physics-informed neural networks (PINNs) have emerged to reconstruct spatial information by using data from sparsely-distributed sensors which are easily collected in real-world applications. However, prevailing approaches use the fully-connected neural network as the backbone without considering the spatial dependency of sensor data, which reduces the accuracy of concentration prediction. This study introduces the physics-informed graph deep learning approach (Physic_GNN) for efficient and accurate hydrogen jet diffusion prediction by using sparsely-distributed sensor data. Graph neural network (GNN) is used to model the spatial dependency of such sensor data by using graph nodes at which governing equations describing the physical law of hydrogen jet diffusion are immediately solved. The computed residuals are then applied to constrain the training process. Public experimental data of hydrogen jet is used to compare the accuracy and efficiency between our proposed approach Physic_GNN and state-of-the-art PINN. The results demonstrate our Physic_GNN exhibits higher accuracy and physical consistency of centerline concentration prediction given sparse concentration compared to PINN and more efficient compared to OpenFOAM. The proposed approach enables accurate and robust real-time spatial consequence reconstruction and underlying physical mechanisms analysis by using sparse sensor data. This facilitates the operation and maintenance management of hydrogen facilities, while concurrently fostering a sustainable transition towards a hydrogen economy.

## 1. Introduction

Rising greenhouse gas emissions have increased global temperatures, prompting efforts to renewable energy sources (RES). A promising climate mitigation strategy is developing a "hydrogen economy" based on green hydrogen production [1,2]. As a clean, versatile fuel for transportation, electricity, heating, and industry, hydrogen demand is projected to expand significantly, potentially reaching 18% of total energy by 2050 [3]. Green hydrogen production using RES, especially solar and wind energy have been extensively explored for large-scale hydrogen production [4,5].

However, one significant challenge associated with hydrogen utilization lies in its propensity to cause embrittlement in metals due to its small atom size, leading to material fatigue and the formation of cracks [6]. This issue is particularly relevant for wind-based and solar-based green-hydrogen facilities, as they are prone to generating cracks or hydrogen bubbles. Inherent to its characteristics, hydrogen can easily leak from infrastructure and disperse into the surrounding environment, forming a potentially flammable vapor cloud. Such a scenario, once ignited, can result in catastrophic fire and explosion disasters [7,8]. Thus, it becomes imperative to develop efficient and accurate spatial concentration prediction models for hydrogen leakage and diffusion to support decision-making processes related to accidental mitigation.

Research progress has been made in terms of hydrogen leakage and diffusion modeling. The prevalent approaches can be classified into three types, namely low-fidelity empirical model, high-fidelity computational fluid dynamics (CFD), and machine/deep learning model. Shu et al., [9] proposed a simplified integral empirical model to predict the concentration of hydrogen buoyant jets with real-time capability. However, the integral model requires empirical constants derived from high-fidelity data to achieve good agreement with experimental observations. Due to the difficulties in conducting experiments on hydrogen leakage, obtaining accurate empirical constants for specific scenarios remains a challenge. CFD has gained significant popularity as it allows for the discretization of the computational domain and enables the numerical solution of differential equations with high accuracy [10,11]. Although high-fidelity CFD models have superior precision, they pose computational demands and are unable to provide real-time prediction of hydrogen leakage and diffusion [12,13].

Machine/deep learning has shown promising results in real-time prediction of gas release and diffusion with remarkable accuracy. Shi et al., [14] introduced a Bayesian regularization artificial neural network (BRANN)-based model, which reliably predicted the volume of flammable cloud. For real-time spatial toxic jets, Na et al., [15] developed a deep learning-based prediction model by integrating a variational autoencoder with a convolutional neural network. Jiao et al., [16] focus on spatial toxic plume prediction, employing deep learning techniques with simulation results from an extensive dataset of 30,022 toxic release scenarios. In a similar vein, Song et al., [17] proposed an integrated encoding-prediction deep neural network that exhibits impressive accuracy for real-time spatiotemporal release concentration prediction. Moreover, Li et al., [6] developed a probabilistic deep learning-based model specifically tailored for real-time hydrogen concentration prediction. While the aforementioned deep learning models demonstrate satisfactory accuracy and real-time prediction capabilities, a key limitation is their reliance on extensive quantities of high-fidelity simulation data for model training. From a practical standpoint, generating the large training datasets required entails prohibitive computational expenses [18]. An additional concern is that neural network predictions may not conform to governing physical laws, presenting the risk of non-physical or unexplainable results [19]. These challenges limit its generalization performance for unseen accidental hydrogen leakage scenarios.

Recently, physics-informed neural networks (PINNs) have emerged as a promising approach to address the challenge of limited data availability and ensure adherence to underlying physical laws. By incorporating prior physical knowledge, PINNs enable learning from small datasets and directly solving Ordinary/Partial Differential Equations (O/PDEs) [20,21]. This involves integrating the solved residuals of O/PDEs into the loss function to optimize the hyper-parameters of the deep learning backbone [22–24]. Notably, Raissi et al., [25] introduced hidden fluid mechanics (HFM), where Navier-Stokes (NS) equations were encoded into the deep neural network through automatic differentiation. In this case, the solved O/PDEs could provide prior physical knowledge of velocity and pressure, effectively constraining the training process and yielding accurate velocity and pressure fields with concentration field for training. Jin et al., [26] also developed physics-informed neural networks for the incompressible NS equations: the velocity-pressure (VP) formulation and the vorticity-velocity (VV) formulation. These formulations enabled the inference of pressure, velocity, and vorticity fields with spatial coordinates. Moreover, Shi et al., [13] integrated physical constraints with deep learning to ensure the physical consistency of spatiotemporal plume boundary concentration prediction. Additionally, Ishitsuka and Lin [27] proposed a PINN that enforces the conservation of mass and energy, ensuring physical plausibility in predicting spatial distributions of temperature, pressure, and permeability of hydrothermal systems. However, despite the notable progress, limited investigations have been conducted to explore the application of PINNs in efficiently and accurately modeling hydrogen leakage and diffusion using sparsely-distributed sensor data as training data. Furthermore, prevalent PINNs employ fully-connected neural networks as the backbone, overlooking the spatial dependency of sensor data, which may reduce the accuracy of spatial concentration prediction [28–30].

This study introduces a physics-informed graph deep learning approach namely Physic_GNN for efficient and accurate modeling of hydrogen diffusion by using sparse sensor data. Graph neural network (GNN) enables the representation of sparse sensor dependencies, with governing equations solved at nodes to constrain training. Public subsonic and under-expanded hydrogen jet data is leveraged for validation versus state-of-the-art PINN. The key contributions of this study are highlighted as follows:

(1) This study pioneers the integration of physical-based graph networks with governing equations for hydrogen diffusion, with validation across low-pressure subsonic and high-pressure under-expanded hydrogen jets.
(2) Physic_GNN demonstrates higher predictive accuracy versus PINN given sparse data and 100x speedup over CFD-based approach OpenFOAM.
(3) This study provides an accurate and efficient alternative for hydrogen spatial consequence analysis and physics elucidation using sparse data, enabling informed real-time operation and maintenance management as hydrogen systems expand.

The remainder of this paper is structured as follows. Section 2 and Section 3 illustrate the mathematical problem and modeling framework of the proposed Physic_GNN. Section 4 provides three validation cases on low-pressure subsonic vertical jets, high-pressure under-expanded vertical jets, and high-pressure under-expanded horizontal jets to demonstrate the effectiveness of Physic_GNN. Finally, discussions and conclusions are provided in section 5 and section 6, respectively.

**2. Problem statement**

The general form of O/PDEs governing the physical law of hydrogen jet diffusion can

be expressed as:

$$\nabla \cdot \mathcal{F}(C(s), \nabla C(s); \boldsymbol{\mu}) = O(C(s), \nabla C(s); \boldsymbol{\mu}) \tag{1}$$

where $s$ is the distance between the interested position and leakage point, $C$ represents concentration, velocity, and other parameters of the jet at s, $\boldsymbol{\mu}$ represents the parameters regarding hydrogen leakage and diffusion scenario, $\nabla$ represents the nonlinear differential operator consisting of the potential solutions of $C(s)$, $\mathcal{F}$ is the flux function, $O$ is the source term. Normally, we would apply numerical solvers such as OpenFOAM, etc., to determine $C(s)$ after collecting experimental concentration data. However, prevalent numerical solutions of $C(s)$ are time-consuming. We thereby employ GNN to immediately solve O/PDEs, which has the potential to reduce the computational cost due to the fast inference capability of neural networks. In addition, GNN could consider the spatial positions and dependency of sensor data, which has the potential to improve the accuracy of spatial hydrogen concentration prediction.

## 3. Proposed approach: physics-informed graph neural network

Fig. 1 illustrates the modeling framework of the proposed Physics-informed graph neural network, namely Physic_GNN, for hydrogen jet diffusion modeling. As can be seen from it, the concentration data from sparsely-distributed sensors is first fed into the input layer. And GNN layers are applied to present spatial positions and dependency of sensors. The latent features from graph layers are fed into the output layer for final prediction. Automatic differentiation (AD) is utilized for differential operation regarding velocity $u$, density $\rho$ etc. Then the residuals namely $\mathcal{L}_{phy}$ generated by solving hydrogen jet governing equations as well as regression loss namely $\mathcal{L}_{re}$ are integrated to determine the optimal hyperparameters such as weights and biases of the deep learning model. The detailed structure of the framework is as follows.

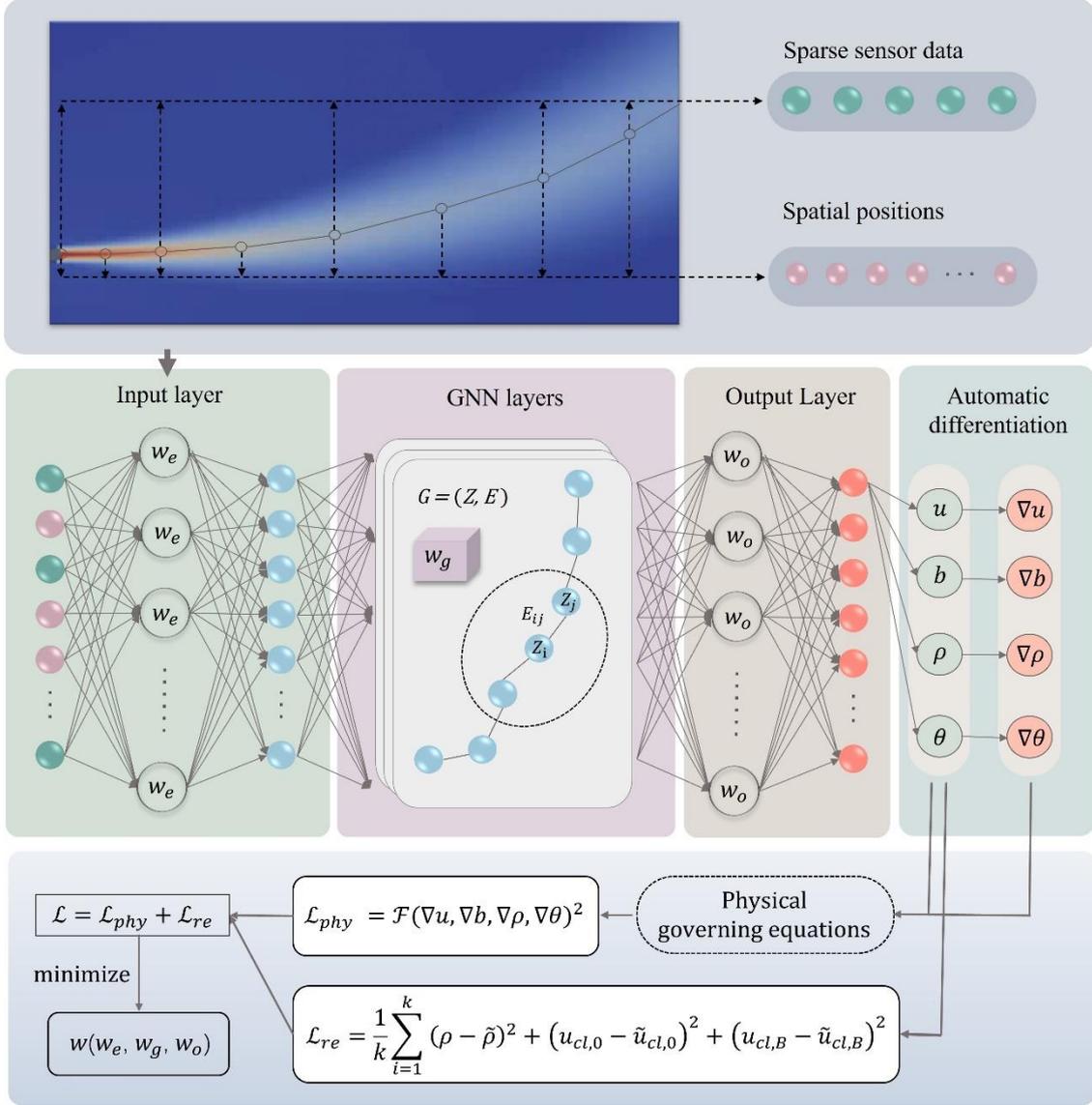

Fig. 1 Architecture of physics-informed graph neural network namely Physic_GNN for hydrogen jet diffusion modeling.

Given sparse observed concentration $\tilde{Y}_{cl} = [\tilde{Y}_{cl,1}, \tilde{Y}_{cl,2}, ..., \tilde{Y}_{cl,k}]$ of $k$ sparse sensor positions $\tilde{S} = [\tilde{S}_1, \tilde{S}_2, ..., \tilde{S}_k]$, the corresponding density can be calculated accordingly, which is expressed as: $\tilde{\rho}_{cl} = [\tilde{\rho}_{cl,1}, \tilde{\rho}_{cl,2}, ..., \tilde{\rho}_{cl,k}]$. The input layer consists of a fully connected network that transforms density $\tilde{\rho}_{cl,i}$ at position $S_i$ to embedding vector $v_i$.

$$v_i = Input\_layer^{w_e}(\tilde{\rho}_{cl,i}) \qquad (2)$$

where $i \in \{1,2,...,k\}$, $w_e$ denotes the parameters in the input layer.

GNN layers are then applied to capture the dependency relationship among spatially-distributed observations, which facilitates concentration prediction of all positions. The association graph structure in GNN denoted as $G=(Z, E)$, is constructed based on node positions. Here, $Z$ represents the nodes corresponding to all the interested positions $S$ ($Z_i = S_i$), and $E$ represents the connections between nodes. The existence of a connection between node $Z_i$ and node $Z_j$ is determined by the adjacency matrix $Ad_{ij}$. By integrating $v_i$ with adjacency matrix $Ad_{ij}$, the latent feature $l_i$ of node $Z_i$ can be obtained via graph network:

$$l_i = GNN^{w_g}(v_i) = Softplus\left(w_g v_i + \sum_{j \in \mathcal{N}(i)} w_g v_j\right) \tag{3}$$

where $Softplus$ represents the activation function, $w_g$ is the weight matrix, and $\mathcal{N}(i)$ represents the set of adjacent nodes of node $Z_i$.

Then, a fully connected network is applied as the output layer to decode latent features to the final prediction, namely velocity $u_i$, width $b_i$, density $\rho_i$, angle of deflection $\theta_i$:

$$C_i(u_i, b_i, \rho_i, \theta_i) = Output\_layer^{w_o}(l_i) \tag{4}$$

where $w_o$ denotes the parameters in the output layer.

Further, four differential operators namely $\nabla u_i$, $\nabla b_i$, $\nabla \rho_i$ and $\nabla \theta_i$, can be approximated by AD in a point-wise manner.

The loss function of the deep neural model is a multi-objective loss function that softly satisfies the physical governing equations of hydrogen jet diffusion and the sparse sensor data. The loss function $\mathcal{L}$ can be expressed as:

$$\mathcal{L} = \mathcal{L}_{phy} + \mathcal{L}_{re} \tag{5}$$

where $\mathcal{L}_{phy}$ represents the residuals derived from physical governing equations of hydrogen jet leakage and diffusion, $\mathcal{L}_{re}$ corresponds to the commonly used regression loss, which aims to achieve a favorable fit to sensor data.

The physical loss $\mathcal{L}_{phy}$ can be expressed as:

$$\mathcal{L}_{phy} = \frac{1}{n}\sum_{i=1}^{n} \mathcal{F}(\nabla u_i, \nabla b_i, \nabla \rho_i, \nabla \theta_i)^2 \tag{6}$$

where $n$ is the number of nodes, the function $\mathcal{F}$ encapsulates the underlying physical processes governing hydrogen jet diffusion, including the continuity equation (Eq.(7)), momentum equation (Eq.(8)), and hydrogen conservation equation (Eq.(9)). Minimizing this loss term during training ensures the model learns parameters consistent with these established physical laws. In effect, the known physics acts as a regularizer on the deep neural network training. By encoding these inductive biases into the loss function, Physic_GNN leverages prior domain knowledge to constrain the solution and achieve physically plausible predictions from limited data.

$$\frac{\partial \rho}{\partial t} + \nabla \cdot (\rho \vec{u}) = 0 \tag{7}$$

where $\vec{u}$ is a tensor of space and time.

$$\frac{\partial (\rho \vec{u})}{\partial t} + \nabla \cdot (\rho \vec{u}\vec{u}) = \rho \vec{g} - \nabla P + \nabla \cdot \tau \tag{8}$$

where $\tau$ is the viscous force and $P$ is the pressure.

$$\frac{\partial Y}{\partial t} + \nabla \cdot (Y\vec{u}) = \nabla \cdot (D\nabla Y) \tag{9}$$

where $D$ is the molecular diffusion coefficient.

The regression loss $\mathcal{L}_{re}$ can be expressed as:

$$\mathcal{L}_{re} = \frac{1}{k}\sum_{i=1}^{k} (\rho_i - \tilde{\rho}_i)^2 + (u_0 - \tilde{u}_0)^2 + (u_B - \tilde{u}_B)^2 \tag{10}$$

where $\tilde{u}_0$ is initial velocity and $\tilde{u}_B$ donates boundary velocity.

By combining these two loss terms, the network can effectively capture the physical behavior of hydrogen jet diffusion while achieving a satisfactory data fit. This approach enables the training of deep learning models with a small number of training samples, thereby enhancing its ability to make accurate predictions beyond the training data.

## 4. Validation and comparison

To demonstrate the effectiveness of Physic_GNN, three validation case studies were performed on low-pressure subsonic vertical jets, high-pressure under-expanded vertical jets, and high-pressure under-expanded horizontal jets. Publicly available experimental data from references [31,32] is leveraged as the benchmark to compare centerline hydrogen concentration and velocity predictions between Physic_GNN and state-of-the-art PINN. Table. 1 summarizes the experimental configurations for the hydrogen leakage scenarios from [31,32] that are modeled in this work.

Table. 1 Experimental configuration of hydrogen leakage and diffusion scenarios.

| Experimental scenario | Pressure in vessel (bar) | Leakage nozzle diameter (mm) | Leakage flow (kg/s) | Jet velocity at the nozzle (m/s) | Jet density at the nozzle (kg/$m^3$) | Froude number |
|---|---|---|---|---|---|---|
| Subsonic vertical jet | - | 1.905 | $6.29 \times 10^{-5}$ | 263.1 | 0.0838 | 527.5 |
| Under-expanded vertical jet | 10 | 1 | - | 1196.3 | 0.521 | 3665.02 |
| Under-expanded horizontal jet | 10 | 3 | 0.045 | 1147 | - | - |

For model development and evaluation, 5 of the 20 experimental concentration values are selected as the training data for Physic_GNN and PINN. Both approaches are implemented in Python 3.6 using TensorFlow 1.14.0 on a system with Intel(R) Xeon(R) Silver 4214R CPU @ 2.40GHz 2.39 GHz and 4 NVIDIA GeForce RTX 2080Ti GPUs, trained for 10,000 epochs with 30 neurons. The developed models are then used to predict centerline concentrations and velocities at the 20 experimental positions. Prediction accuracy is quantified by calculating the mean squared error (MSE) between the predicted and experimentally measured concentrations across the 20 evaluation points. This rigorous framework enabled a direct comparison of the concentration and velocity prediction performance of the proposed Physic_GNN versus state-of-the-art PINN given limited training data.

**4.1 low-pressure subsonic vertical hydrogen jets**

4.1.1 Governing equations of low-pressure subsonic vertical hydrogen jets

The physical theory of subsonic free jet is employed to describe hydrogen leakage and diffusion. The dimensionless parameter, namely the exit densitometric Froude number $Fr_{den}$ is used to model the proportional relationship between momentum and buoyancy force during hydrogen leakage and diffusion [33]. The exit densitometric Froude number $Fr_{den}$ can be expressed as:

$$Fr_{den} = \frac{u_{exit}}{\left(\frac{gd_e|\rho_{exit} - \rho_\infty|}{\rho_{exit}}\right)^{1/2}} \quad (11)$$

where $u_{exit}$ is the nozzle exit velocity of the jet, $g$ is the acceleration of gravity, $d_e$ is the leakage diameter, $\rho_\infty$ is the ambient air density, and $\rho_{exit}$ is the gas density at the leakage point under the specific pressure and temperature in the vessel. Depending on different $Fr_{den}$ values, hydrogen jet can be categorized into three types, namely plume ($Fr_{den}$ less than 10), buoyancy-dominated jet ($Fr_{den}$ ranging from 10 to 1000) and momentum-dominated jet ($Fr_{den}$ larger than 1000) [34].

Hydrogen leaking from a low-pressure vessel into the atmosphere is typically dominated by initial momentum and buoyancy, namely buoyancy-dominated jets. A typical subsonic vertical hydrogen jet can be shown in Fig. 2. In the figure, $u_0$ is the initial jet velocity, $u_{cl}$ is velocity of the jet on the central axis, $Y_0$ is the initial concentration, $Y$ is concentration of the jet, $Y_{cl}$ is concentration of the jet on the central axis, $s$ represents the coordinate along the centerline of the jet, and the direction of gravity is parallel to the direction of the *S*-axis and straight down.

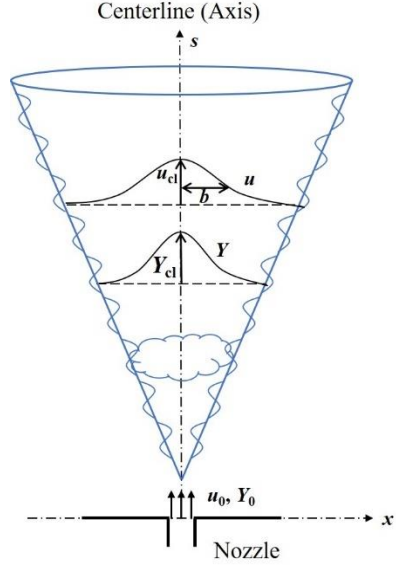

Fig. 2 Schematic of subsonic vertical hydrogen jet.

The assumptions made in the jet model are as follows:

(1) Pressure and Temperature Equality: The pressure and temperature of the jet at the leakage point are considered equivalent to those of the ambient air. This assumption implies that heat exchange between hydrogen and air can be neglected.

(2) Quasi-One-Dimensional Flow: The radial velocity component of the jet is assumed to be negligibly small compared to the axial velocity (along the jet centerline). As a result, the flow is approximated as quasi-one-dimensional.

(3) Quasi-Steady Flow: The jet flow is assumed to exhibit a quasi-steady behavior. This assumption implies that the time-averaged quantities of flow properties, such as velocity, density, and concentration values, do not vary with time.

(4) Gaussian Distribution Profiles: The local axial velocity, density, and mass fraction of the jet are assumed to follow Gaussian distribution profiles. This assumption allows for a simplified representation of the spatial distribution of these properties along the jet centerline [35].

The assumptions can be expressed as:

$$u(s,r) = u_{cl}(s)e^{-\frac{r^2}{b^2}} \qquad (12)$$

$$\rho_\infty - \rho(s,r) = (\rho_\infty - \rho_{cl}(s))e^{-\frac{r^2}{\lambda^2 b^2}} \qquad (13)$$

$$\rho(s,r)Y(s,r) = \rho_{cl}(s)Y_{cl}(s)e^{-\frac{r^2}{\gamma^2 b^2}} \qquad (14)$$

where $r$ is radial distance, $\rho_\infty$ is density of the ambient air, $\rho_{cl}$ is density of the jet on the central axis, $\lambda$ is the buoyancy dispersion coefficient, which expresses the difference in spreading rates between the velocity and density deficiency profiles [36]. The mass fraction of the jet on the central axis $Y_{cl}$ can be expressed as:

$$Y_{cl} = \frac{M_{H_2}}{M_{air} - M_{H_2}} \left( \frac{\rho_\infty}{\rho_{cl}} - 1 \right) \tag{15}$$

where, $M_{H_2}$ is the molecular mass of hydrogen, $M_{air}$ is the molecular mass of air.

Thus, the integral form of the continuity equation, namely Eq.(7), can be expressed as:

$$\frac{\partial}{\partial s} \int_0^\infty \rho u r \, dr = \rho_\infty E \tag{16}$$

where $E$ is the entrainment rate of the jet to air. The entrainment of the jet to air for a subsonic jet can be expressed as [37]:

$$E = E_{mom} + E_{buoy} \tag{17}$$

where $E_{mom}$ represents entrainment caused by momentum, and $E_{buoy}$ represents entrainment caused by buoyancy. The momentum entrainment $E_{mom}$ can be calculated by [38]:

$$E_{mom} = \beta_A \left( \frac{\pi d_e^2}{4} \frac{\rho_0 u_0^2}{\rho_0} \right)^{1/2} \tag{18}$$

where $\beta_A$ represents empirical coefficient.

The buoyancy entrainment $E_{buoy}$ can be calculated as:

$$E_{buoy} = \frac{a_2}{Fr_1} (2\pi u_{cl}) b \tag{19}$$

where $Fr_1 = \frac{u_{cl}^2}{gb \frac{\rho_\infty - \rho_{cl}}{\rho_0}}$, $a_2$ is an empirical constant determined by the density Froude number at the jet outlet:

$$a_2 = \begin{cases} 17.313 - 0.11665 * Fr_{den} + 2.0771 \times 10^{-4} * Fr_{den}^2, & Fr_{den} < 268 \\ 0.97, & Fr_{den} \geq 268 \end{cases} \tag{20}$$

Besides, the entrainment $E$ also can be expressed as:

$$E = 2\pi b \alpha u_{cl} \tag{21}$$

where $\alpha$ is entraining constant. As the buoyancy effect becomes more and more significant, a large number of experimental results show that the entraining constant approaches a limit value of 0.082 [39]. Therefore, in the calculation, firstly, the entrainment $E$ is calculated by Eq. (17), and then the empirical constant α is calculated by Eq. (21). If the calculated value is greater than 0.082, the value is set to 0.082, and $E$ is recalculated by Eq. (21).

The integral form of momentum equation, namely Eq.(8), can be expressed as:

$$\frac{\partial}{\partial s}\int_0^\infty \rho u^2 r dr = \int_0^\infty (\rho_\infty - \rho) g r dr \tag{22}$$

The integral form of hydrogen conservation equation, namely Eq.(9), can be expressed as:

$$\frac{\partial}{\partial s}\int_0^\infty \rho u(Y - Y_\infty) r dr = 0 \tag{23}$$

where $Y_\infty$ is the hydrogen mass fraction in the ambient air ($Y_\infty = 0$).

Substituting Eq. (17) - (21) into Eq. (16), Eq. (22), and Eq. (23), and integrating $r$ from 0 to positive infinity, three ODEs with distance $s$ on the center line as the independent variable are obtained:

$$b^2\big(\rho_\infty - \Lambda_1(\rho_\infty - \rho_{cl})\big)\frac{du_{cl}}{ds} + 2u_{cl}b\big(\rho_\infty - \Lambda_1(\rho_\infty - \rho_{cl})\big)\frac{db}{ds} + u_{cl}b^2\Lambda_1\frac{d\rho_{cl}}{ds} = \frac{\rho_\infty E}{\pi} \tag{24}$$

$$\left(1 - \Lambda_2\frac{(\rho_\infty - \rho_{cl})}{\rho_\infty}\right)\left(u_{cl}b^2\frac{du_{cl}}{ds} + u_{cl}^2 b\frac{db}{ds}\right) + u_{cl}^2\frac{b^2}{2}\Lambda_2\frac{1}{\rho_\infty}\frac{d\rho_{cl}}{ds} = \frac{(\rho_\infty - \rho_{cl})}{\rho_\infty}g\lambda^2 b^2 \tag{25}$$

$$\rho_{cl} y_{cl} b^2\frac{du_{cl}}{ds} + 2u_{cl}\rho_{cl}y_{cl}b\frac{db}{ds} + u_{cl}b^2\left(y_{cl} - \frac{M_{H_2}M_{air}}{M_{air} - M_{H_2}}\frac{p}{RT\rho_{cl}}\right)\frac{d\rho_{cl}}{ds} = 0 \tag{26}$$

where $\Lambda_1 = \frac{\lambda^2}{1+\lambda^2}$, $\Lambda_2 = \frac{2\lambda^2}{1+2\lambda^2}$.

4.1.2 Prediction of low-pressure subsonic vertical hydrogen jets

*MSE* values between the predicted 20 concentration values and experimental values are calculated, as shown in Table. 2. From it, one may see that our proposed Physic_GNN achieves a lower *MSE* value. This indicates that Physic_GNN predicts the concentration values from sparse observations more accurately.

Table. 2 *MSE* between the predicted 20 concentration values and experimental values of the subsonic vertical hydrogen jet.

| Experimental scenario | MSE | |
|---|---|---|
| | Physic_GNN | PINN |
| Subsonic vertical hydrogen jet | 2.77E-03 | 12.82 |

Fig. 3 illustrates the predicted centerline concentration curves as a function of dimensionless axial distance by our proposed Physic_GNN and PINN. Please noting that the dimensionless axial distance is defined as the ratio of distance $S$ between the centerline point and leakage nozzle to the diameter of leakage nozzle $d_e$. As can be seen, the predicted hydrogen concentration curves by our proposed Physic_GNN exhibit a hyperbolic decay trend along the diffusion centerline. However, it is observed that the concentration values predicted by PINN deviate significantly from the experimental measurements. Relatively, the concentration values predicted by our proposed Physic_GNN align well with the experimental measurements.

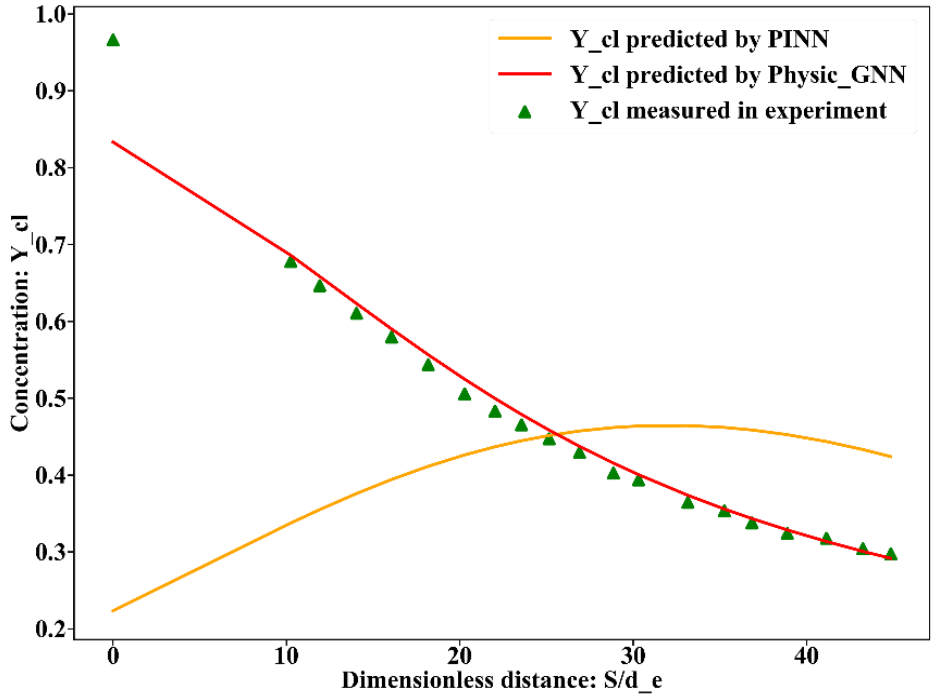

Fig. 3 Prediction centerline concentration curves of subsonic vertical hydrogen jet as a function of dimensionless axial distance by our proposed Physic_GNN and PINN

Fig. 4 shows the predicted centerline velocity curves as a function of dimensionless axial distance by our proposed Physic_GNN and PINN, benchmarked against the velocity field generated by OpenFOAM under an identical subsonic vertical hydrogen jet diffusion scenario. While the velocities predicted by both Physic_GNN and PINN are higher than that predicted by OpenFOAM, it can be observed that PINN yields values markedly exceeding those from the OpenFOAM simulation across the domain. In contrast, the velocity curve predicted by Physic_GNN demonstrates closer agreement with the OpenFOAM benchmark.

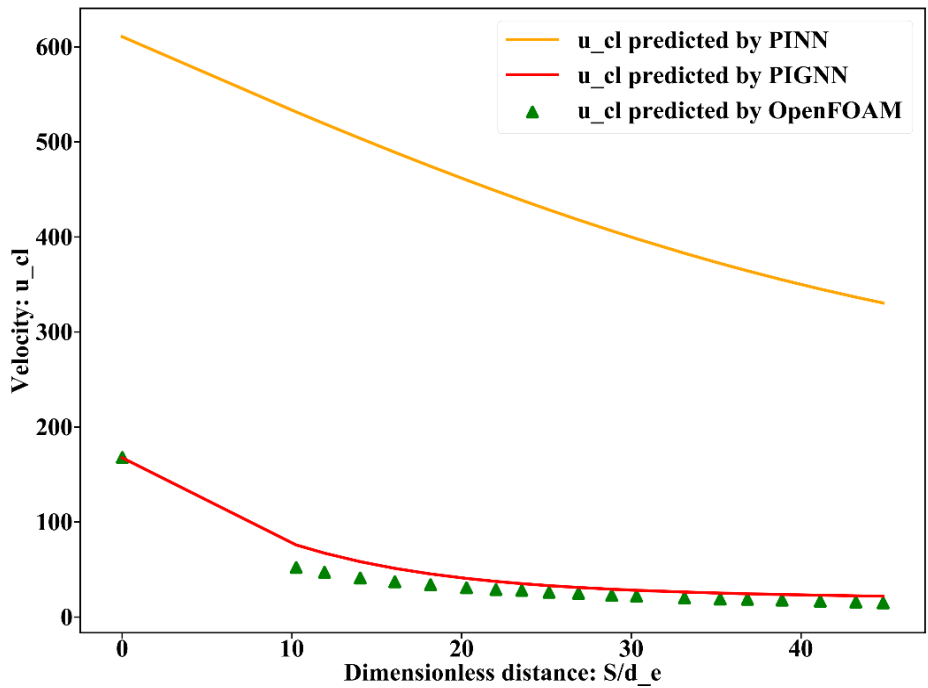

Fig. 4 Prediction centerline velocity curves of subsonic vertical hydrogen jet as a function of dimensionless axial distance by our proposed Physic_GNN and PINN

Table. 3 demonstrates the comparison of computational efficiency between our proposed Physic_GNN, PINN, and OpenFOAM. Please noting the OpenFOAM simulation is performed by using with computer server of Intel(R) Xeon(R) Silver 4214R CPU @ 2.40GHz 2.39 GHz without GPU configuration. From it, one may see both our proposed Physic_GNN and PINN could predict the centerline hydrogen concentration in less than 2 minutes, which is more computational efficiency compared to OpenFOAM which requires almost 4 hours to get the predictions with desirable accuracy. All the above findings demonstrate that our Physic_GNN exhibits higher accuracy of centerline concentration prediction given sparsely experimental concentration compared to PINN and more efficient compared to OpenFOAM.

Table. 3 Computational efficiency of our proposed approach, PINN, and OpenFOAM

| Approach | Inference time/s |
|---|---|
| Physic_GNN | 90 |
| PINN | 120 |
| OpenFOAM | 14400 |

**4.2 high-pressure under-expanded vertical hydrogen jets**

4.2.1 Governing equations of high-pressure under-expanded vertical hydrogen jets

When the high-pressure stored hydrogen gas is released to the ambient atmosphere through a small orifice, a high-pressure under-expanded jet is created. This jet expands rapidly over a short distance outside the orifice, forming a complex surge structure known as the Mach disk [40]. After passage through the Mach disk, the pressure in the core of the jet increases to ambient pressure and the velocity decreases to subsonic levels. For this reason, the subsonic jet model is not directly applicable to high-pressure under-expanded jets, and the notional nozzle model is required to reduce complex under-expanded jets to subsonic jets.

In this paper, the virtual nozzle model proposed by Birch [41,42] is introduced to calculate the "pseudo-diameter" at the outlet of the virtual nozzle, and then the parameters at the virtual outlet are used as the inlet conditions of the subsonic jet equations. Fig. 5 shows a schematic diagram of the virtual nozzle model, where $d_v$ represents the calculated pseudo-diameter and the three different gas states represent the stagnation state in the high-pressure vessel (Level 0), the sound velocity flow state at the leakage orifice (Level 1), and the state at the outlet of the virtual nozzle (Level 2).

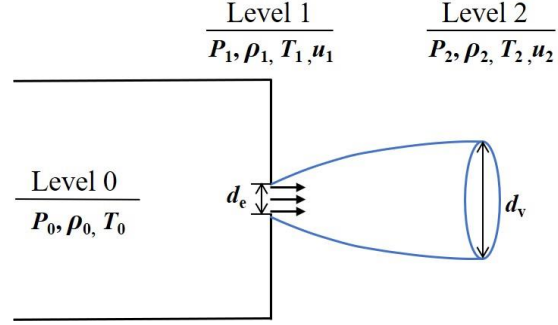

Fig. 5 Schematic of notional nozzle model.

Assuming that the viscous forces are negligible over the expansion surface, and there is no entrainment of ambient fluid, then the equations for conservation of mass and momentum can be written as:

$$\rho_1 u_1 A_1 = \rho_2 u_2 A_2 \tag{27}$$

where $A_1$ represents the cross-sectional area of the actual leakage orifice and $A_2$ represents the cross-sectional area of the virtual nozzle.

The momentum conservation equation for the flow from the leakage (LEVEL 1) to the virtual nozzle (LEVEL 2) is as follows:

$$\rho_2 u_2^2 A_2 = \rho_1 u_1^2 A_1 + (P_1 - P_2) A_1 \tag{28}$$

Assuming $P_2$ equals the ambient pressure $P_\infty$, the velocity $u_2$ and area $A_2$ of the virtual nozzle can be expressed as:

$$u_2 = u_1 + \frac{P_1 - P_2}{\rho_1 u_1} \tag{29}$$

$$A_2 = \frac{\rho_1^2 u_1^2 A_1}{\rho_2 (P_1 - P_\infty + \rho_1 u_1^2)} \tag{30}$$

Compared with the enthalpy of the airflow, the heat transfer and friction loss between the wall surface and the airflow can be ignored, that is, the flow between the stagnation state (LEVEL 0) and the leakage orifice (LEVEL 1) is assumed to be an adiabatic flow, so the airflow parameters at the leakage can be calculated according to the entropy-relation formula:

$$P_e = P_1 = P_0 \left(\frac{2}{\gamma + 1}\right)^{\frac{\gamma}{\gamma - 1}} \tag{31}$$

$$\rho_e = \rho_1 = P_0 \left(\frac{2}{\gamma + 1}\right)^{\frac{1}{\gamma - 1}} \frac{M_{H_2}}{R T_1} \tag{32}$$

$$u_e = u_1 = \sqrt{\frac{RT_1}{M_{H_2}}(\frac{2\gamma}{\gamma+1})} \tag{33}$$

where $\gamma$ is the specific heat ratio of the leaking gas.

Then, according to the ideal gas equation of state, the airflow density at the leakage outlet is calculated as follows:

$$\rho_e = \rho_1 = \frac{p_0 M_{H_2}}{RT_0}\left(\frac{2}{\gamma+1}\right)^{\frac{1}{\gamma-1}} \tag{34}$$

Thus, the pseudo-diameter $d_v$ can be calculated by:

$$\frac{d_v}{d_e} = \frac{\rho_1 u_1}{\sqrt{\rho_{H_2}(P_1 - P_\infty + \rho_1 u_1^2)}} \tag{35}$$

The flow parameters of the high-pressure jet can be calculated by inputting the virtual nozzle into the subsonic jet equations.

4.2.2 Prediction of high-pressure under-expanded vertical hydrogen jets

Table 4 shows the *MSE* values for the predicted high-pressure under-expanded vertical hydrogen jets obtained from PINN and Physic_GNN. From it, one may see the *MSE* values obtained by our proposed Physic_GNN model are smaller than that by PINN. This indicates that Physic_GNN provides more accurate predictions of the centerline concentration for the high-pressure under-expanded jet scenario compared to PINN.

Table. 4 MSE between the predicted 20 concentration values and experimental values of the high-pressure under-expanded vertical hydrogen jet.

| Experimental scenario | *MSE* | |
|---|---|---|
| | Physic_GNN | PINN |
| Under-expanded vertical hydrogen jet | 4.17E-03 | 15.58 |

Fig. 6 depicts the centerline concentration predictions as a function of dimensionless axial distance for the high-pressure under-expanded vertical hydrogen jet diffusion. It can be observed that the concentration curve predicted by Physic_GNN shows increasingly closer agreement with the experimental measurements as the dimensionless distance increases. In contrast, the PINN model yields concentration values exhibiting an upward trend with increasing dimensionless distance, contrary to physical principles governing jet diffusion. Overall, the improved conformity of the Physic_GNN results to experimental data highlights its superior ability to capture the true physical behavior of jet diffusion compared to PINN.

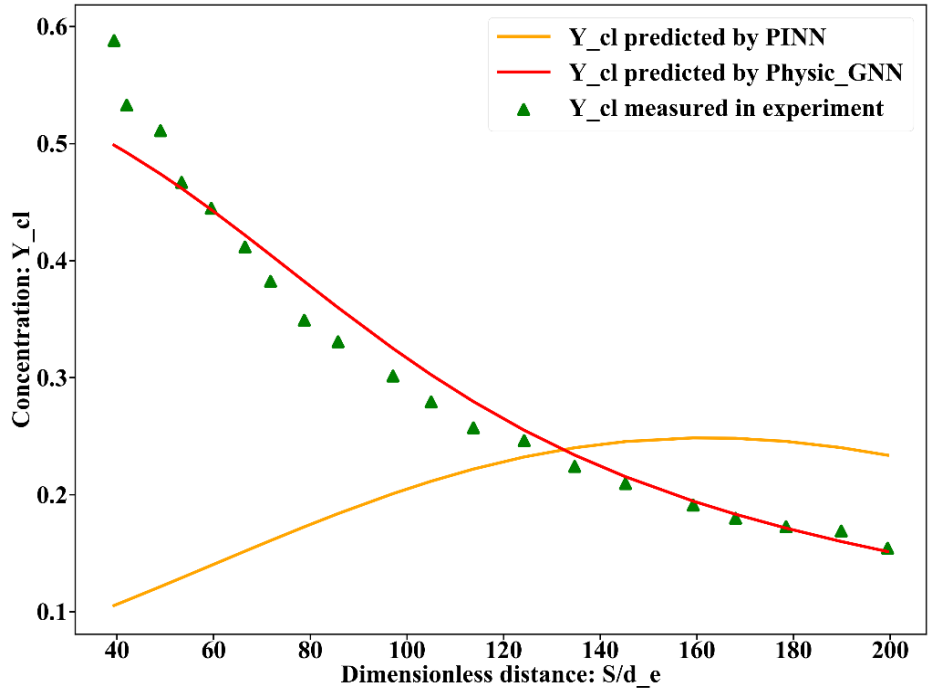

Fig. 6 Centerline concentration prediction of an under-expanded vertical hydrogen jet.

Fig. 7 shows the predicted centerline hydrogen velocity curves by our proposed Physic_GNN and PINN, compared against the velocity generated by using OpenFOAM under the same configuration of the experimental subsonic jet diffuse scenario. It can be seen that while the velocity values predicted by both PINN and Physic_GNN exhibit the expected hyperbolic decay trend along the centerline, the PINN model yields velocity values markedly higher than those from the OpenFOAM simulation. In contrast, the velocity profile predicted by Physic_GNN shows much closer agreement with the OpenFOAM result.

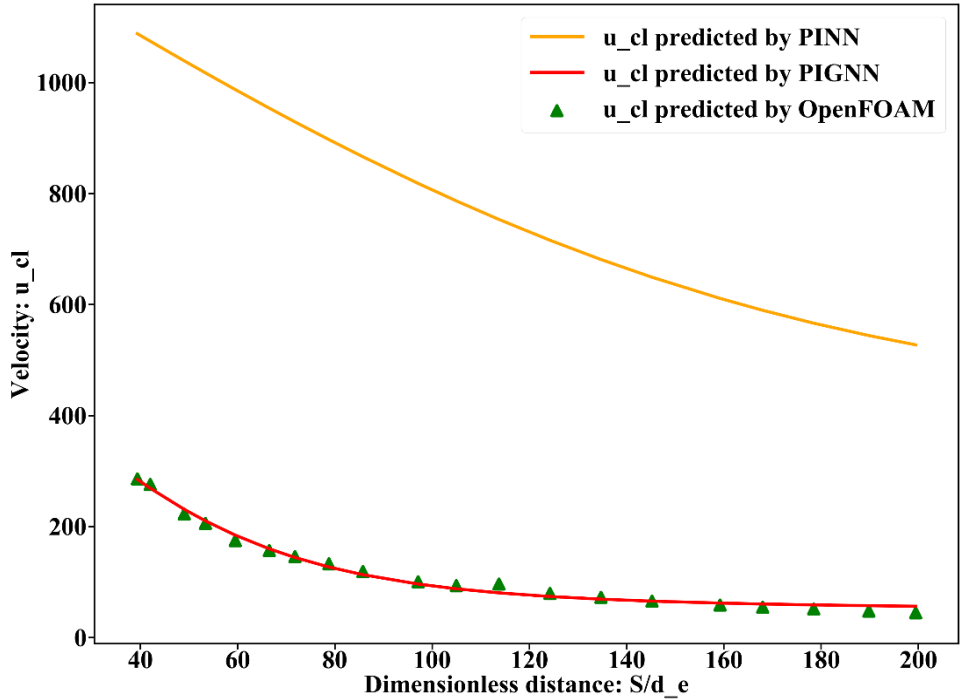

Fig. 7 Prediction centerline velocity curves of under-expanded vertical hydrogen jet as a function of dimensionless axial distance by our proposed Physic_GNN and PINN

A comparison of the computational efficiency for modeling the under-expanded vertical jet scenario, over a longer predicted distance than the subsonic case, is presented in Table 5 for Physic_GNN, PINN, and OpenFOAM. It can be observed that while the prediction times increase for all three methods compared to the subsonic jet modeling, Physic_GNN and PINN are still able to generate results in approximately 2 minutes. In contrast, the required computational time for OpenFOAM rises substantially, to nearly 10 hours. These results highlight that even for more complex under-expanded jet dynamics, the proposed Physic_GNN approach maintains a significant efficiency advantage over traditional computational fluid dynamics simulations.

Table. 5 Computational efficiency of our proposed approach, PINN, and OpenFOAM

| Approach | Inference time/s |
| --- | --- |
| Physic_GNN | 96 |
| PINN | 124 |
| OpenFOAM | 36000 |

## 4.3 high pressure under-expanded horizontal hydrogen jets

4.3.1 Governing equations of high pressure under-expanded horizontal hydrogen jet

For the high-pressure under-expanded horizontal hydrogen jet, the virtual nozzle at the outlet is first calculated using Eq. 3 to account for the expansion. This virtual nozzle is then utilized as the inlet of subsonic horizontal jets. Fig. 8 shows the schematic of a horizontal hydrogen jet, where the Angle between the S-axis and the horizontal direction is $\theta$, and the azimuth angle along the cross-section of the jet is $\varphi$.

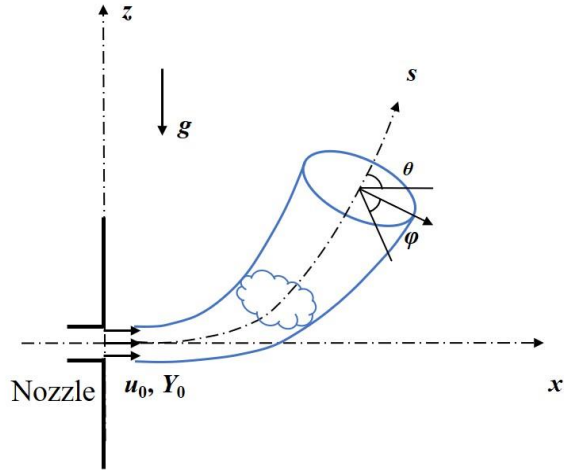

Fig. 8 Schematic of subsonic horizontal hydrogen jet.

Correspondingly, the continuity equation can be expressed as:

$$\frac{\partial}{\partial s}\int_0^{2\pi}\int_0^{\infty} \rho u r dr d\varphi = \rho_\infty E \tag{36}$$

The momentum equation in the x-direction can be expressed as:

$$\frac{\partial}{\partial s}\int_0^{2\pi}\int_0^{\infty} \rho u^2 cos\theta r dr d\varphi = 0 \tag{37}$$

The momentum equation in the z-direction can be expressed as:

$$\frac{\partial}{\partial s}\int_0^{2\pi}\int_0^{\infty} \rho u^2 sin\theta r dr d\varphi = \int_0^{2\pi}\int_0^{\infty} (\rho_\infty - \rho) g r dr d\varphi \tag{38}$$

The hydrogen conservation equation can be expressed as:

$$\frac{\partial}{\partial s}\int_0^{2\pi}\int_0^{\infty} \rho u (Y - Y_\infty) r dr d\varphi = 0 \tag{39}$$

In addition, the coordinate equation of the jet center line can be expressed as:

$$\frac{\partial}{\partial s}(x) = cos\theta \tag{40}$$

$$\frac{\partial}{\partial s}(z) = \sin\theta \tag{41}$$

Taking Eq. (12)- (15) into Eq.(36) - Eq.(41) and integrating $r$ from 0 to positive infinity and $\varphi$ from 0 to $2\pi$, four differential equations with distance $s$ on the centerline are obtained:

$$b^2(\rho_\infty - \Lambda_1(\rho_\infty - \rho_{cl}))\frac{du_{cl}}{ds} + 2u_{cl}b(\rho_\infty - \Lambda_1(\rho_\infty - \rho_{cl}))\frac{db}{ds} + u_{cl}b^2\Lambda_1\frac{d\rho_{cl}(s)}{ds} = \frac{\rho_\infty E}{\pi} \tag{42}$$

$$\left(1 - \Lambda_2\frac{(\rho_c - \rho_{cl})}{\rho_\infty}\right)\left(u_{cl}\cos\theta b^2\frac{du_{cl}}{ds} - u_{cl}^2\sin\theta\frac{b^2}{2}\frac{d\theta}{ds} + u_{cl}^2\cos\theta b\frac{db}{ds}\right) + u_{cl}^2\cos\theta\frac{b^2}{2}\Lambda_2\frac{1}{\rho_\infty}\frac{d\rho_{cl}}{ds} = 0 \tag{43}$$

$$\left(1 - \Lambda_2\frac{(\rho_\infty - \rho_{cl})}{\rho_\infty}\right)\left(u_{cl}\sin\theta b^2\frac{du_{cl}}{ds} + u_{cl}^2\cos\theta\frac{b^2}{2}\frac{d\theta}{ds} + u_{cl}^2\sin\theta b\frac{db}{ds}\right) + u_{cl}^2\sin\theta\frac{b^2}{2}\Lambda_2\frac{1}{\rho_\infty}\frac{d\rho_{cl}}{ds} = \frac{(\rho_\infty - \rho_{cl})}{\rho_\infty}g\lambda^2 b^2 \tag{44}$$

$$\rho_{cl}y_{cl}b^2\frac{du_{cl}}{ds} + 2u_{cl}\rho_{cl}y_{cl}b\frac{db}{ds} + u_{cl}b^2\left(y_{cl} - \frac{M_{H_2}M_{air}}{M_{air} - M_{H_2}}\frac{p}{RT\rho_{cl}}\right)\frac{d\rho_{cl}}{ds} = 0 \tag{45}$$

4.2.2 Prediction of high pressure under-expanded horizontal hydrogen jets

It should be noted that the 20 concentration values of high pressure under-expanded horizontal hydrogen jet for the training of Physic_GNN and PINN models are obtained from OpenFOAM simulations. This approach is taken because the number of experimental measurements available is less than 20, and prior work by Keenan et al., [43] validated the ability of OpenFOAM to accurately model the concentration field for high-pressure horizontal hydrogen release against experiments conducted by Shell and the UK Health and Safety Laboratory [32].

The *MSE* values achieved by Physic_GNN and PINN for predicting the centerline jet concentrations are presented in Table 6. It can be observed that Physic_GNN attains substantially a lower *MSE* compared to PINN.

Table. 6 *MSE* between the predicted 20 concentration values and experimental values of high-pressure under-expanded horizontal hydrogen jets.

| Experimental scenario | MSE | |
|---|---|---|
| | Physic_GNN | PINN |
| Under-expanded horizontal hydrogen jet | 6.35E-03 | 6.73 |

Fig. 9 presents the predicted concentration values as a function of dimensionless distance for the high-pressure horizontal hydrogen jet scenario obtained from Physic_GNN, PINN, and OpenFOAM. Please noting that the figure also shows the measured concentration values in the experiment. It can be observed that the concentration values predicted by Physic_GNN show better agreement with the OpenFOAM benchmark compared to PINN. Although the Physic_GNN concentrations are slightly higher than OpenFOAM in the region $z/d_e<500$, the overall decay trend is consistent with experimental measurements and the established physics of jet diffusion. In contrast, the PINN model yields an increasing concentration with dimensionless distance, contrary to the expected attenuation behavior.

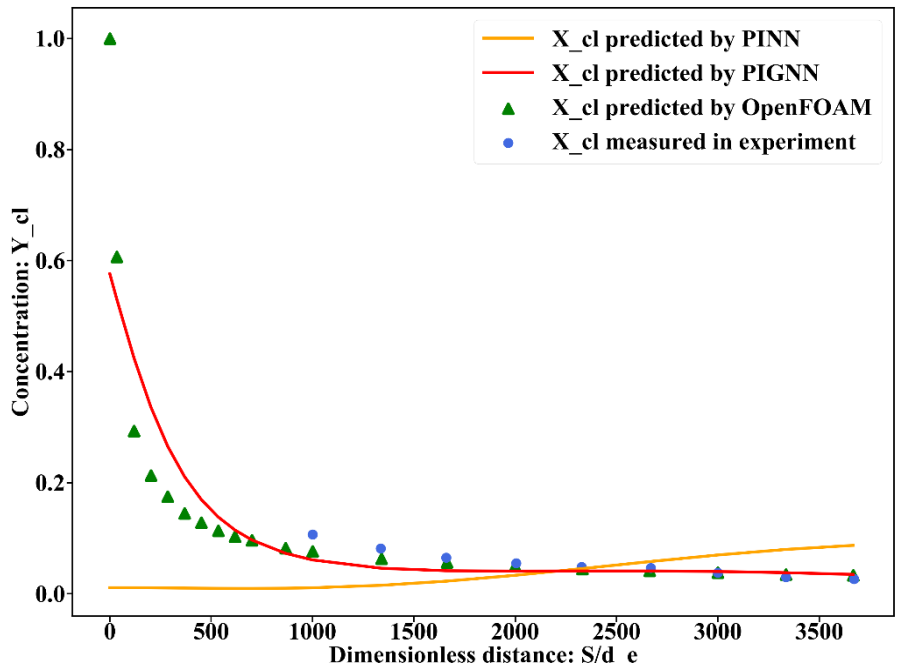

Fig. 9 Centerline concentration prediction of an under-expanded horizontal hydrogen jet.

Fig. 10 depicts the centerline hydrogen velocity curves by the proposed Physic_GNN and PINN models as a function of dimensionless axial distance. As can be seen, the velocity values and variation trend predicted by PINN differ markedly from the OpenFOAM results. In contrast, the velocities predicted by Physic_GNN are in close agreement with that predicted by OpenFOAM and exhibit the expected decay trend with increasing dimensionless distance, consistent with established physics governing jet velocity attenuation.

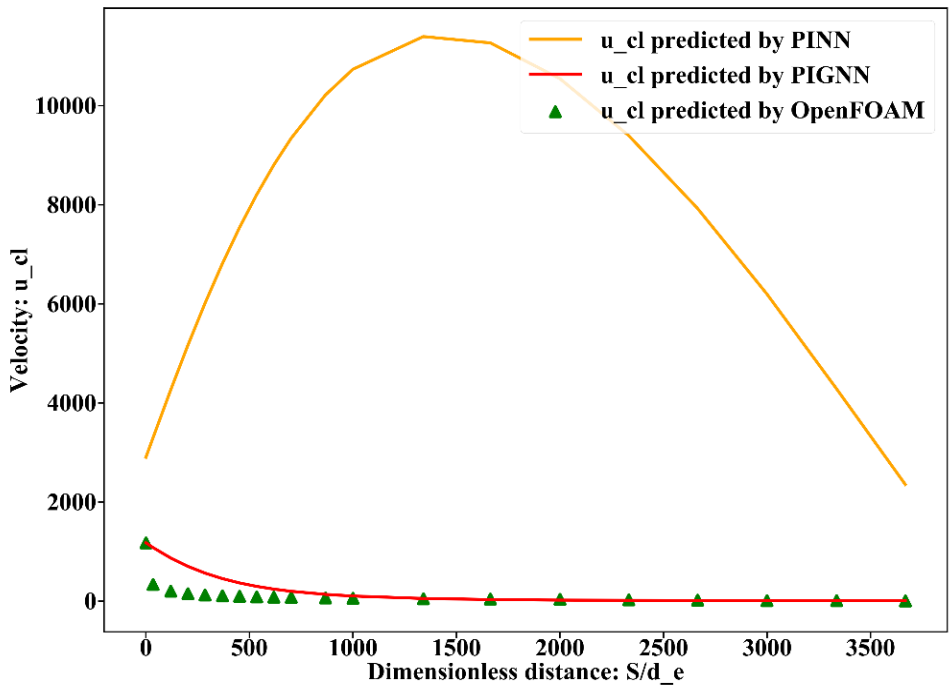

Fig. 10 Centerline concentration prediction of an under-expanded horizontal hydrogen jet.

Table 7 presents the computational efficiency comparison of Physic_GNN, PINN, and OpenFOAM. It can be observed that Physic_GNN and PINN demonstrate significantly lower prediction times, with Physic_GNN generating results in under 2 minutes. In contrast, the OpenFOAM simulations require approximately 1 day of computation time to converge.

Table. 7 Computational efficiency of our proposed approach, PINN, and OpenFOAM

| Approach | Inference time/s |
|---|---|
| Physic_GNN | 116 |
| PINN | 150 |
| OpenFOAM | 86400 |

## 5. Discussions

This study provides a novel approach for efficient and accurate modeling of hydrogen leakage and diffusion by using sparsely-distributed sensor data. Compared with PINNs, our proposed Physic_GNN model exhibits enhanced prediction accuracy when dealing with limited sensor data. Moreover, in contrast to computationally intensive CFD-based approaches like OpenFOAM, our model demonstrates superior time efficiency.

The application of Physic_GNN enables real-time spatial distribution prediction of subsonic and under-expanded hydrogen jets based on sparse sensor data. This capability can significantly contribute to making informed operation and maintenance strategy in wind-based and solar-based green-hydrogen production. The method's versatility also allows for its implementation in operation and maintenance management across hydrogen production, storage, and utilization processes, with potential implications to provide a robust support for constructing a digital twin of hydrogen industry.

Additionally, our proposed method allows for the reconstruction of concentration and velocity fields that adhere to physical laws based on experimental measurements. This valuable feature empowers researchers to gain deeper insights into the physical mechanisms underlying hydrogen leakage and diffusion observed in experiments.

In conclusion, this study introduces a cutting-edge approach to addressing critical challenges in hydrogen leakage and diffusion modeling, providing an accurate and efficient data-driven alternative to support safe, sustainable management as hydrogen systems scale up.

Future works are expected to enhance the proposed physics-informed graph neural network by integrating momentum conservation equations, such as the compressible NS equations, to enable predictions of concentration and velocity fields in more complex two-dimensional and three-dimensional scenarios involving unintended hydrogen releases.

6. Conclusions

This work presented Physic_GNN, a physics-informed graph neural network for hydrogen leakage modeling from sparse sensor data. Publicly available subsonic and under-expanded jet data validated the superior accuracy and efficiency of our model. The key conclusions are:

(1) Physic_GNN demonstrates higher accuracy and physical consistency in predicting centerline concentrations for both subsonic and under-expanded jets from limited sensors versus PINN.

(2) Prediction times under 2 minutes are achieved, enabling 100x speedup over computational fluid dynamics.

(3) This work encoded hydrogen physics in graph deep learning for real-time spatial consequence reconstruction and physical mechanisms elucidation using sparse data, facilitating the operation and maintenance of green hydrogen production.

7. Acknowledgements

This study was supported by National Key R&D Program of China [grant number 2021YFB4000901-03]. National Natural Science Foundation of China (Project No.: 52101341). Natural Science Foundation of Shandong Province (Project No.: ZR2020KF018). China Postdoctoral Science Foundation Funded Project (Project No.: 2019M662469). Qingdao Science and Technology Plan (Project No.: 203412nsh). Key Project of Natural Science Foundation of Shandong Province (Project No.: ZR2020KF018). The authors would like to acknowledge partial support of the Hong Kong Research Grants Council (T22-505/19-N).

**References**

[1]   Superchi F, Mati A, Carcasci C, Bianchini A. Techno-economic analysis of wind-powered green hydrogen production to facilitate the decarbonization of hard-to-abate sectors: A case study on steelmaking. Appl Energy

2023;342:121198. https://doi.org/10.1016/j.apenergy.2023.121198.

[2] Müller LA, Leonard A, Trotter PA, Hirmer S. Green hydrogen production and use in low- and middle-income countries: A least-cost geospatial modelling approach applied to Kenya. Appl Energy 2023;343:121219. https://doi.org/10.1016/j.apenergy.2023.121219.

[3] Wang W, Ma Y, Maroufmashat A, Zhang N, Li J, Xiao X. Optimal design of large-scale solar-aided hydrogen production process via machine learning based optimisation framework. Appl Energy 2022;305:117751. https://doi.org/10.1016/j.apenergy.2021.117751.

[4] Sui J, Chen Z, Wang C, Wang Y, Liu J, Li W. Efficient hydrogen production from solar energy and fossil fuel via water-electrolysis and methane-steam-reforming hybridization. Appl Energy 2020;276:115409. https://doi.org/10.1016/j.apenergy.2020.115409.

[5] Abadie LM, Chamorro JM. Investment in wind-based hydrogen production under economic and physical uncertainties. Appl Energy 2023;337. https://doi.org/10.1016/j.apenergy.2023.120881.

[6] Li J, Xie W, Li H, Qian X, Shi J, Xie Z, et al. Real-time hydrogen release and dispersion modelling of hydrogen refuelling station by using deep learning probability approach. Int J Hydrogen Energy 2023. https://doi.org/10.1016/j.ijhydene.2023.04.126.

[7] Shi J, Xie W, Li J, Zhang X, Huang X, Usmani AS, et al. Real-time plume tracking using transfer learning approach. Comput Chem Eng 2023;172:108172. https://doi.org/10.1016/j.compchemeng.2023.108172.

[8] Shu Z, Liang W, Liu F, Lei G, Zheng X, Qian H. Diffusion characteristics of liquid hydrogen spills in a crossflow field: Prediction model and experiment. Appl Energy 2022;323:119617. https://doi.org/10.1016/j.apenergy.2022.119617.

[9] Shu Z, Liang W, Zheng X, Lei G, Cao P, Dai W, et al. Dispersion characteristics of hydrogen leakage: Comparing the prediction model with the experiment. Energy 2021;236:121420. https://doi.org/10.1016/j.energy.2021.121420.

[10] Boghi A, Di Venuta I, Gori F. Passive scalar diffusion in the near field region of turbulent rectangular submerged free jets. Int J Heat Mass Transf 2017;112:1017–31. https://doi.org/10.1016/j.ijheatmasstransfer.2017.05.038.

[11] Zhou C, Yang Z, Chen G, Zhang Q, Yang Y. Study on leakage and explosion consequence for hydrogen blended natural gas in urban distribution networks. Int J Hydrogen Energy 2022;47:27096–115. https://doi.org/10.1016/j.ijhydene.2022.06.064.

[12] Shi J, Xie W, Huang X, Xiao F, Usmani AS, Khan F, et al. Real-time natural gas release forecasting by using physics-guided deep learning probability model. J Clean Prod 2022;368:133201. https://doi.org/10.1016/j.jclepro.2022.133201.

[13] Shi J, Li J, Usmani AS, Zhu Y, Chen G, Yang D. Probabilistic real-time deep-water natural gas hydrate dispersion modeling by using a novel hybrid deep

learning approach. Energy 2021;219:119572. https://doi.org/10.1016/j.energy.2020.119572.

[14] Shi J, Khan F, Zhu Y, Li J, Chen G. Robust data-driven model to study dispersion of vapor cloud in offshore facility. Ocean Eng 2018;161:98–110. https://doi.org/10.1016/j.oceaneng.2018.04.098.

[15] Na J, Jeon K, Lee WB. Toxic gas release modeling for real-time analysis using variational autoencoder with convolutional neural networks. Chem Eng Sci 2018;181:68–78. https://doi.org/10.1016/j.ces.2018.02.008.

[16] Jiao Z, Ji C, Sun Y, Hong Y, Wang Q. Deep learning based quantitative property-consequence relationship (QPCR) models for toxic dispersion prediction. Process Saf Environ Prot 2021;152:352–60. https://doi.org/10.1016/j.psep.2021.06.019.

[17] Song D, Lee K, Phark C, Jung S. Spatiotemporal and layout-adaptive prediction of leak gas dispersion by encoding-prediction neural network. Process Saf Environ Prot 2021;151:365–72. https://doi.org/10.1016/j.psep.2021.05.021.

[18] Tu H, Moura S, Wang Y, Fang H. Integrating physics-based modeling with machine learning for lithium-ion batteries. Appl Energy 2023;329:120289. https://doi.org/10.1016/j.apenergy.2022.120289.

[19] Di Natale L, Svetozarevic B, Heer P, Jones CN. Towards scalable physically consistent neural networks: An application to data-driven multi-zone thermal building models. Appl Energy 2023;340. https://doi.org/10.1016/j.apenergy.2023.121071.

[20] Bünning F, Huber B, Schalbetter A, Aboudonia A, Hudoba de Badyn M, Heer P, et al. Physics-informed linear regression is competitive with two Machine Learning methods in residential building MPC. Appl Energy 2022;310:118491. https://doi.org/10.1016/j.apenergy.2021.118491.

[21] Gokhale G, Claessens B, Develder C. Physics informed neural networks for control oriented thermal modeling of buildings. Appl Energy 2022;314:118852. https://doi.org/10.1016/j.apenergy.2022.118852.

[22] Zhang J, Zhao X. Spatiotemporal wind field prediction based on physics-informed deep learning and LIDAR measurements. Appl Energy 2021;288:116641. https://doi.org/10.1016/j.apenergy.2021.116641.

[23] Zhang J, Zhao X. Three-dimensional spatiotemporal wind field reconstruction based on physics-informed deep learning. Appl Energy 2021;300:117390. https://doi.org/10.1016/j.apenergy.2021.117390.

[24] Yin X, Wen K, Huang W, Luo Y, Ding Y, Gong J, et al. A high-accuracy online transient simulation framework of natural gas pipeline network by integrating physics-based and data-driven methods. Appl Energy 2023;333:120615. https://doi.org/10.1016/j.apenergy.2022.120615.

[25] Raissi M, Yazdani A, Karniadakis GE. Hidden fluid mechanics: Learning velocity and pressure fields from flow visualizations. Science (80- ) 2020;367:1026–30.

[26] Jin X, Cai S, Li H, Karniadakis GE. NSFnets (Navier-Stokes flow nets):

Physics-informed neural networks for the incompressible Navier-Stokes equations. J Comput Phys 2021;426:109951. https://doi.org/10.1016/j.jcp.2020.109951.

[27] Ishitsuka K, Lin W. Physics-informed neural network for inverse modeling of natural-state geothermal systems. Appl Energy 2023;337:120855. https://doi.org/10.1016/j.apenergy.2023.120855.

[28] Jiang L, Wang L, Chu X, Xiao Y, Zhang H. PhyGNNet: Solving spatiotemporal PDEs with Physics-informed Graph Neural Network. Proc. 2023 2nd Asia Conf. Algorithms, Comput. Mach. Learn., 2022, p. 1–7.

[29] Gao H, Zahr MJ, Wang JX. Physics-informed graph neural Galerkin networks: A unified framework for solving PDE-governed forward and inverse problems. Comput Methods Appl Mech Eng 2022;390:114502. https://doi.org/10.1016/j.cma.2021.114502.

[30] Pagnier L, Chertkov M. Physics-Informed Graphical Neural Network for Parameter & State Estimations in Power Systems 2021:1–12.

[31] Li X, Hecht ES, Christopher DM. Validation of a reduced-order jet model for subsonic and underexpanded hydrogen jets. Int J Hydrogen Energy 2016;41:1348–58. https://doi.org/10.1016/j.ijhydene.2015.10.071.

[32] Roberts PT, Shirvill LC, Roberts TA, Butler CJ, Royle M. Dispersion of hydrogen from high-pressure sources. Inst Chem Eng Symp Ser 2006:410–21.

[33] Giannissi SG, Venetsanos AG, Hecht ES. Numerical predictions of cryogenic hydrogen vertical jets. Int J Hydrogen Energy 2021;46:12566–76. https://doi.org/10.1016/j.ijhydene.2020.08.021.

[34] Yang F, Wang T, Deng X, Dang J, Huang Z, Hu S, et al. Review on hydrogen safety issues: Incident statistics, hydrogen diffusion, and detonation process. Int J Hydrogen Energy 2021;46:31467–88. https://doi.org/10.1016/j.ijhydene.2021.07.005.

[35] Schefer RW, Houf WG, Williams TC. Investigation of small-scale unintended releases of hydrogen: Buoyancy effects. Int J Hydrogen Energy 2008;33:4702–12. https://doi.org/10.1016/j.ijhydene.2008.05.091.

[36] El-Amin MF, Kanayama H. Similarity consideration of the buoyant jet resulting from hydrogen leakage. Int J Hydrogen Energy 2009;34:5803–9. https://doi.org/10.1016/j.ijhydene.2009.05.059.

[37] Houf W, Schefer R. Analytical and experimental investigation of small-scale unintended releases of hydrogen. Int J Hydrogen Energy 2008;33:1435–44. https://doi.org/10.1016/j.ijhydene.2007.11.031.

[38] Ricou FP, Spalding DB. Measurements of entrainment by axisymmetrical turbulent jets. J Fluid Mech 1961;11:21–32. https://doi.org/10.1017/S0022112061000834.

[39] List EJ, Imberger J. Turbulent entrainment in buoyant jets and plumes. J Hydraul Div 1973;99:1461–74.

[40] Jugroot M, Groth CPT, Thomson BA, Baranov V, Collings BA. Numerical investigation of interface region flows in mass spectrometers: Neutral gas transport. J Phys D Appl Phys 2004;37:1289–300.

https://doi.org/10.1088/0022-3727/37/8/019.

[41] Birch AD, Brown DR, Dodson MG, Swaffield F. The structure and concentration decay of high pressure jets of natural gas. Combust Sci Technol 1984;36:249–61. https://doi.org/10.1080/00102208408923739.

[42] Birch A, Hughes D, Swaffield F. Velocity Decay of High Pressure Jets. Combust Sci Technol 1987;52:161–71.

[43] Keenan JJ, Makarov D V., Molkov V V. Modelling and simulation of high-pressure hydrogen jets using notional nozzle theory and open source code OpenFOAM. Int J Hydrogen Energy 2017;42:7447–56. https://doi.org/10.1016/j.ijhydene.2016.07.022.